\documentclass[twocolumn,superscriptaddress,showpacs,preprintnumbers,amsmath,amssymb,nofootinbib]{revtex4-2}
\usepackage{graphicx}
\usepackage[normalem]{ulem}
\usepackage{textcomp}   
\usepackage[scaled]{helvet}  
\usepackage{amsmath,amssymb}  
\usepackage{type1cm}
\usepackage{bm}  
\usepackage{color}

\begin{document}

\title{Fermion Casimir effect and magnetic Larkin-Ovchinnikov phases}

\author{Antonino Flachi} 
\affiliation{Department of Physics,  Keio University, 4-1-1 Hiyoshi, Kanagawa 223-8521, Japan}
\affiliation{Research and Education Center for Natural Sciences, Keio University, 4-1-1 Hiyoshi, Kanagawa 223-8521, Japan}

\author{Muneto Nitta} 
\affiliation{Department of Physics,  Keio University, 4-1-1 Hiyoshi, Kanagawa 223-8521, Japan}
\affiliation{Research and Education Center for Natural Sciences, Keio University, 4-1-1 Hiyoshi, Kanagawa 223-8521, Japan}
\affiliation{International Institute for Sustainability with Knotted Chiral Meta Matter (WPI-SKCM2), Hiroshima University, Higashi-Hiroshima, Hiroshima 739-8526, Japan}

\author{Satoshi Takada} 
\affiliation{Department of Mechanical Systems Engineering and Institute of Engineering, Tokyo University of Agriculture and Technology, 2–24–16 Naka-cho, Koganei, Tokyo 184–8588, Japan}

\author{Ryosuke Yoshii} 
\affiliation{International Institute for Sustainability with Knotted Chiral Meta Matter (WPI-SKCM2), Hiroshima University, Higashi-Hiroshima, Hiroshima 739-8526, Japan}
\affiliation{Center for Liberal Arts and Sciences, Sanyo-Onoda City University, Yamaguchi 756-0884, Japan}
\affiliation{Research and Education Center for Natural Sciences, Keio University, 4-1-1 Hiyoshi, Kanagawa 223-8521, Japan}
\date{\today}

\begin{abstract}
This paper explores how magnetic fields affect the Casimir effect within the context of a simple quasi-1D interacting fermionic system. A novel phenomenon emerges, resulting from the interaction between external magnetic fields and boundary conditions, which alters the ground state in complex ways and leads to first-order phase transitions among various ground states, specifically inhomogeneous solutions with differing node counts. We calculate the Casimir force, which 
exhibits
discontinuities (jumps) at the transition points between the different ground states. 
\end{abstract}

\maketitle

\section{Introduction} 

A fundamental prediction of quantum physics is the presence of random fluctuations throughout the vacuum. When this concept is applied to the electromagnetic field, it results in a minute force between uncharged boundaries that arises from the alteration of the spectrum of zero-point energies of electromagnetic fluctuations due to changes in spatial topology created by boundary conditions. This phenomenon is known as the Casimir effect \cite{Casimir:1948dh}. Since its theoretical proposal in 1948 and the definitive observations made in the late 1990s \cite{Experiments}, the Casimir effect has been the focus of extensive theoretical and experimental research, confirming its occurrence in various setups, including metals, semiconductors, magnetic materials, superconductors, liquid crystals, vacuums, and liquids and gases, as well as across different boundary shapes (See \cite{Experiments,textbook,magnetic,metamaterials,Mund,PhysRevLett.121.030405,PhysRevA.99.052507,Munday,Dant,Chernodub:2019,Chernodub:2013} for examples of related work). This vigorous exploration has been primarily driven by the desire to gain a deeper understanding of quantum vacuum effects at a fundamental level and by potential applications in nanotechnology. The field has now entered a phase of applications, ranging from micro and nano electro-mechanical systems (MEMS/NEMS) (e.g., \cite{Capasso1}) and electronic components like transistors and diodes, leveraging quantum vacuum effects to the realm of technology (see Refs.~\cite{Xu:2022nn,Xu:2022nc,Zhang:2024} for some recent examples).

An area of recent attention has been that of superconductors. When the boundaries superconduct, constraints on the quantum fluctuations occur across the critical temperature $T_c$ due to a shift in the optical properties of the superconductor, resulting in a modification of the Casimir force. In spite of the effect being small and difficult to measure (since the superconducting transition only alters the frequencies of order $\sim k_B T_c$), successful experiments have been carried out. A particularly interesting realization is the on-chip platform which uses as boundaries two 
superconducting parallel strings, one of which is attached to a movable mirror of an opto-mechanical cavity, whose resonance frequency depends on the distance between the strings and can be measured \cite{PhysRevLett.121.030405}. 
A theoretical framework has been worked out in Ref. \cite{PhysRevA.99.052507}, which also discusses some related theoretical ideas and experiments. Examples of other relevant work on the Casimir effect for free or interacting fermions can be found in Refs.~\cite{CasFer1,CasFer2,CasFer3,CasFer4,CasFer5,CasFer6,CasFer7,CasFer8,CasFer9,CasFer10,CasFer11,CasFer12,CasFer13,CasFer14}.

The concept of superconductors as catalysts for intriguing quantum vacuum phenomena remains rich with opportunities for exploration. This, as in the previous example, is partly due to the unique optical properties of superconductors, but also due to the fact that superconductivity requires interactions. This latter aspect has been discussed in relation to the Casimir effect with the underlying idea being the fact that the quantum fluctuations occur over an inhomogeneous ground state whose shape and features depend on the strength of the interactions \cite{Flachi:2012pf,Flachi:2013bc,Flachi:2017cdo}. In this work, we will investigate a new direction in the interplay between superconductivity and quantum vacuum effects in the presence of an external field. To be specific, the primary focus will be on the fermion fluctuations occurring in a superconductor subjected to a magnetic field, a setup physically significant due to two types of phenomena. 

The first relates to changes in the fermion condensate that occurs at the onset of the superconducting or superfluid transition. At the critical point, Cooper pairs form as a result of the instability of the Fermi sea against attractive interactions, leading to the formation of a non-vanishing condensate $\langle \bar \psi \psi \rangle$. This transition alters the characteristics of fluctuations, allowing them to behave as massive and consequently affecting the vacuum energy. Such changes can be influenced by thermal effects or variations in interaction strength, as noted in earlier studies. 

The second and more intriguing aspect involves the influence of an external magnetic field. (See Refs.~\cite{{Zhang:2024,Chernodub:2013,Cougo-Pinto:1999,Metalidis:2002,Erdas:2013jga,Sitenko:2014,Nakata:2023}} for earlier work related to the Casimir effect in the presence of magnetic fields.)
In general, Cooper pair formation is affected by magnetic fields due to the spin imbalance caused by Zeeman splitting. This spin imbalance results in an excess of spins which cannot accommodate pairing and thus costs energy. For magnetic fields larger than a critical value, the superconducting phase falls into the normal phase. Besides the normal and superconducting phases, in the presence of a magnetic field other configurations in which the condensate $\langle \bar \psi\psi\rangle$ has nodes, may become energetically favorable. 
In these phases, known as Larkin-Ovchinnikov (LO) phases \cite{Larkin:1964wok}, any spin in excess can be accommodated around the nodes and become energetically preferred to homogeneous configurations \cite{Machida:1984zz,Yoshii:2011yt}. 
Evidence of the LO phase is observed for a superfluid in 
cold atom systems where the spin imbalance is directly induced  \cite{PMID:20882011}. Such modulated ground states with nodes were also studied in the context of QCD with possible applications to neutron star interiors \cite{Casalbuoni:2003wh,Anglani:2013gfu,Buballa:2014tba}.

In this paper, we will show, within a simple and tractable model, that the Casimir energy due to the fermion fluctuations not only depends on the number of nodes in the ground state and can be controlled by changes in the size of the system and by the intensity of the magnetic field, but it develops discontinuities (singularities) that correspond to first order phase transitions. 
The results we shall present below indicate that, for the present setup, a jump in the Casimir energy could be evidence of a first order transition between different inhomogeneous ground states.

\section{The model}

Our starting point will be a systems of fermions with attractive interaction that we describe with the following model 
\begin{equation}
\mathcal{L}=\bar\psi i\gamma^\mu \partial_\mu \psi+g^2 (\bar \psi \psi)^2. 
\label{lagrangian} 
\end{equation}
This model is adopted in both condensed matter physics (known as the BCS model) 
or in high energy physics (known as the Gross-Neveu model). 
Here $\psi$ represents the spinor field, $\gamma^\mu$ the gamma matrices, 
$\bar \psi=\psi^\dagger \gamma^0 $, and $g^2$ is the coupling strength. 
We implicitly assume that the model derives from a 2D model by some unspecified mechanism of compactification to quasi-1D. This may happen due the presence of a confining potential that effectively reduces the dimensionality. The model will be coupled to a magnetic field that appears as a constant shift to the chemical potential of the system. We note that the quasi-1D setup is also relevant to analyze the experimental system examined in Ref.~\cite{PMID:20882011}.
In the mean field approximation, this setup is known to lead to the following set of (discretized) Bogoliubov-de Gennes coupled equations
\begin{align}
&\sum_{j} \left[
\begin{array}{cc}
H_{i,j,\sigma}&\Delta_i^\ast \delta_{i,j}\\
\Delta_i^\ast \delta_{i,j}&-H_{i,j,\bar\sigma}
\end{array}
\right]\left[
\begin{array}{cc}
u^{(n)}_{j,\sigma}\\
v^{(n)}_{j,\bar\sigma}
\end{array}
\right]=E_n\left[
\begin{array}{cc}
u^{(n)}_{i,\sigma}\\
v^{(n)}_{i,\bar\sigma}
\end{array}
\right],\label{BdG}\\
& \Delta_i=g^2\sum_n u^{(n)}_{j,\uparrow}v^{(n)\ast}_{j,\downarrow}\tanh{\frac{E_n}{2T}},
\label{Gap}
\end{align}
where $H_{i,j,\sigma}=t\delta_{i,j}-(\mu+\sigma h)\delta_{i,j}$ and $\sigma\bar\sigma=-1$. The quantities $u_{i,\sigma}$ and $v_{i,\sigma}$ represent the eigenstates (two-spinor components) at sites labelled by $i$ ($1\le i\le N$). The system is assumed to have size $L=Na$ with $a$ being the lattice constant. In the above equations, $t$, $\mu$, and $T$ stand for the transfer integral, the chemical potential, and the temperature, respectively; $h$ is the magnetic (Zeeman) field, which causes a spin-dependent energy splitting $h\sigma$ ($\sigma=\pm 1$). Switching off the magnetic field gives the standard mean field BdG equations for a system of fermions with a four-point interactions. 

As explained in the previous section, the calculations try to uncover the type of modifications that can be induced in the Casimir energy when a magnetic field is present. This is not a trivial generalization. A magnetic field causes a spin-dependent energy splitting  and allows for the formation of energetically favored inhomogeneous solutions.

The crucial point of our argument is that the shape of the ground state is expected to change with the magnetic field: a change in the topology of the ground state, due to continuity and the quasi-one dimensionality, is expected to take place as a change in the number of nodes of the condensate. Moreover, the change cannot be continuous, thus leaving us with the intuition that a discontinuous, first-order phase transition should affect the ground state. It is then natural to expect that similar discontinuous transitions should occur in the Casimir energy that originate from the fluctuations occurring on top of the ground state. The goal of the paper is to examine this idea and show whether and how such discontinuities eventually take place.

\section{Numerical solutions}

The target of the numerical calculation is to minimize the free energy as a function of the ``vacuum'' (ground state) solutions obtained from Eqs.\ \eqref{BdG} and \eqref{Gap}, and thus obtaining the lowest energy state. The free energy $F(h,L)$ in the mean field approximation can be expressed as 
\begin{align}
F(h,L) = E_{GS} + E_{PC} + E_{T},
\label{freeen}
\end{align}
where the first term,
\begin{align}
E_{GS} = \sum_i\left({\Delta_i^2}/{2g^2}-\mu-h\right),
\end{align}
is the ground-state (condensate) contribution, the quantum vacuum energy term is
\begin{align}
E_{PC} = \sum_{n=1}^{2N} E_n,
\end{align}
that we call ``pseudo-Casimir term'' (this will be explained more precisely later), plus a thermal correction,
\begin{align}
E_{T} = - T\sum_{n=1}^{2N} \ln \left(1+e^{E_n/T}\right).
\end{align}
The second term returns the usual fermionic Casimir energy in the limit $g\to 0$, $a\to 0$ ($N\to \infty$) in absence of both chemical potential and magnetic field, while the first background term vanishes in the same limit. The last term survives and gives the thermal correction to the Casimir energy. The same problem, at zero temperature and for vanishing magnetic field was discussed in Ref.~\cite{Flachi:2017cdo}.

The system of equations (\ref{BdG}), (\ref{Gap}), (\ref{freeen}) can be solved numerically in a self-consistent manner by starting from a trial-function for $\Delta$ as initial input, looking for numerical solutions for the two-spinor eigenstates, recalculating $\Delta$ and repeat until convergence is met. 

\begin{figure}[h]
    \centering
    \includegraphics[width=\linewidth]{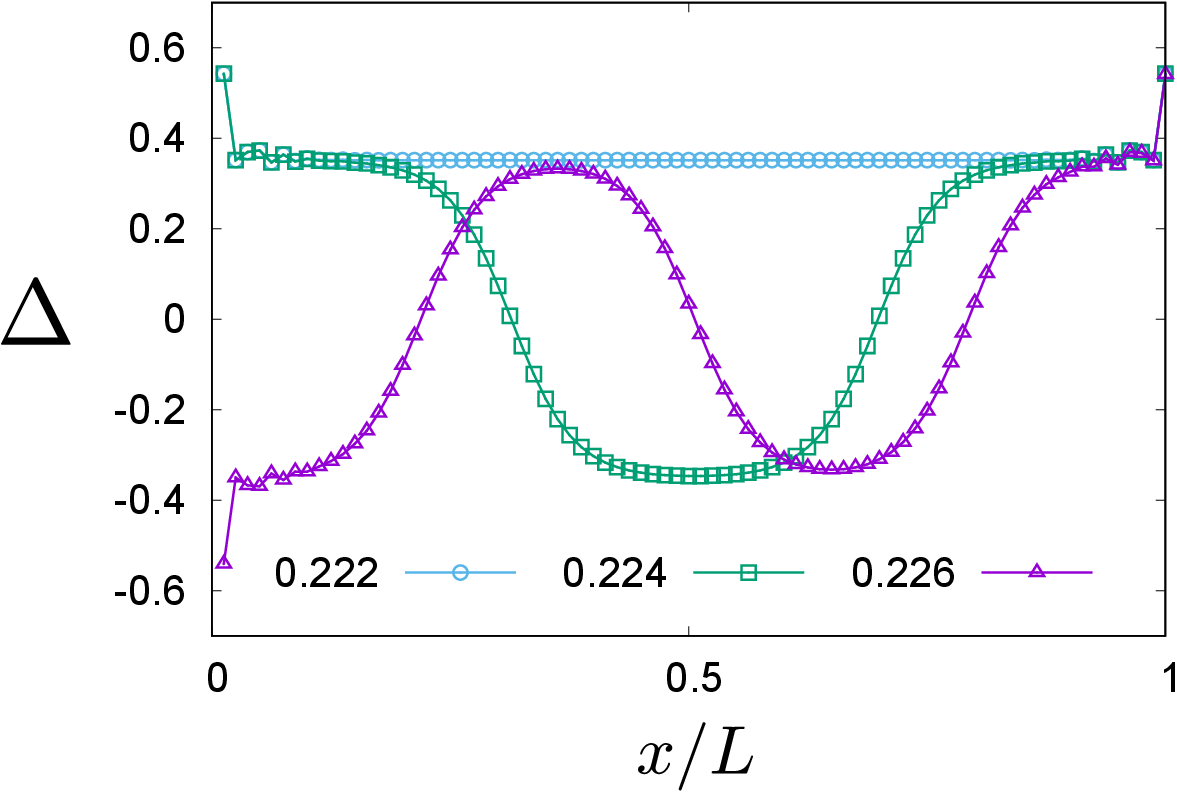}
    \caption{Configuration of $\Delta(x)$ for the lowest energy state for $h=0.25$. The size of the system is set to be $L/a=80$. The inhomogeneous solution with the six nodes obtained by the numerics. The condensate oscillates in the middle region and becomes large at the vicinities of boundaries.} 
    \label{Delta1}
\end{figure}

A set of illustrative solutions for the ground state are shown in Fig.~\ref{Delta1}, where we plot $\Delta$  as a function of the normalized spatial coordinate $x/L$ for different combinations of parameters yielding a ground state with different number of nodes (the corresponding numerical values of the parameters are: $g/t=1$, $\mu/t=-0.5$, and $T/t=0.005$, and $h/t=0.222, 0.224$, and $0.226$, respectively for the solutions with zero, two and three nodes). The near-boundary divergence of $\Delta$ originates from the requirement that the spinor solutions are regular at the boundary when Dirichlet boundary conditions are imposed. This can be seen by expanding the spinor solutions $u$ and $v$ and noticing that the $v$ solution is regular at the boundaries only when $\Delta$ diverges. This property is not peculiar to the presence of a magnetic field, but is a generic consequence of the BdG equation (\ref{BdG}) on an interval \cite{Flachi:2017cdo} when ideal sharp boundaries are considered. While near the boundaries the solution departs from the usual LO solution, for a system large enough, it reaquires the features of the LO phases away from the boundaries. 

\begin{figure}
    \centering
    \includegraphics[width=\linewidth]{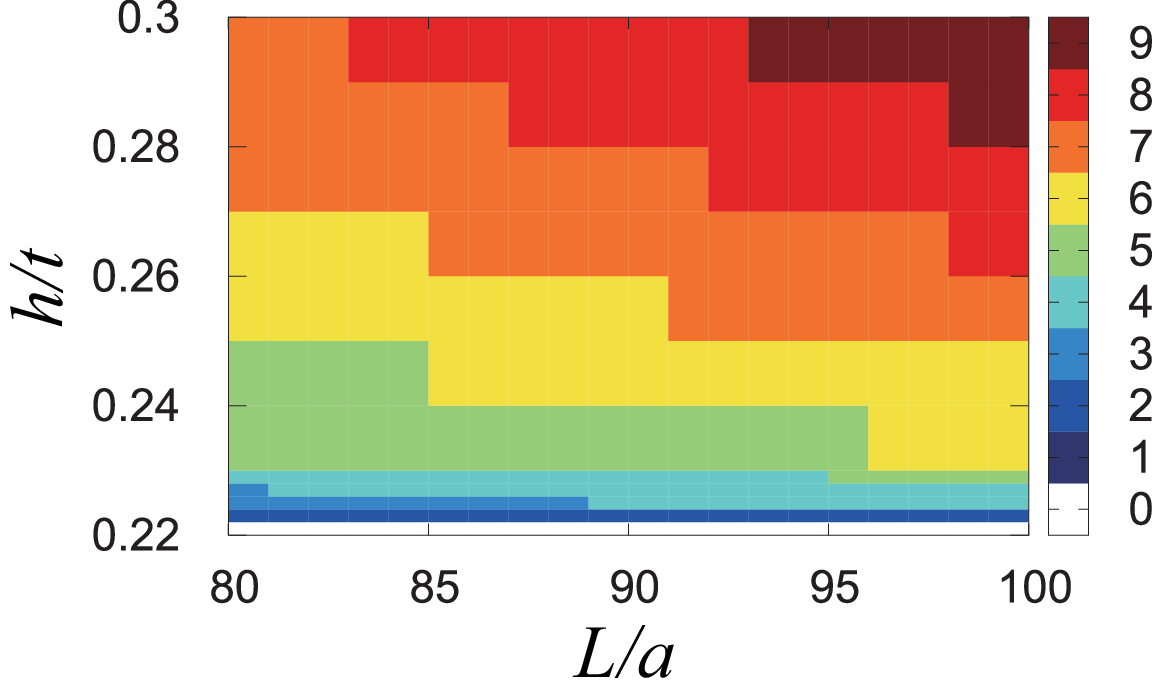}
    \caption{Phase diagram in the $L$-$h$ plane. The physical parameters as in Fig.~\ref{Delta1} and choice of the range in the figure has been chosen for illustration. The numbers in the figure refer to the number of nodes of the order parameter in the lowest energy state. All the transitions are first order transition. Here we set the length scale and the energy scale to be $a$ and $t$, respectively. 
    }
    \label{PhaseDiagram}
\end{figure}

Fig.~\ref{Delta1} basically conveys the idea that a magnetic field causes the number of nodes in the ground state to change. Since the number of nodes is a discrete quantity, then we expect the transition between different solutions as a function of the magnetic field to change discontinuously. This can be explicitly seen by computing the free energy as a function of $L$ and $h$, minimizing it with respect to $\Delta$ and counting the number of nodes. This phase diagram (in the $L-h$ plane) is displayed in Fig.\ \ref{PhaseDiagram}. The first evident feature is that increasing the size of the system $L$ or the magnetic field $h$ corresponds to an increase in the number of nodes in the ground state configuration, consistently with the earlier argument on number of nodes vs excess spins. Similarly, since the number of nodes is an integer which cannot change continuously, the transition between different solutions should be discontinuous (first-order). Fig.\ \ref{PhaseDiagram} also reveals that a smaller number of nodes in the ground state is favored for small $L$: starting from any given ground state configuration and reducing $L$ while keeping the number of nodes fixed is energetically costly, as the background has to become steeper around the nodes; it is energetically more advantageous if the number of nodes decreases.

\begin{figure}
    \centering
    \includegraphics[width=\linewidth]{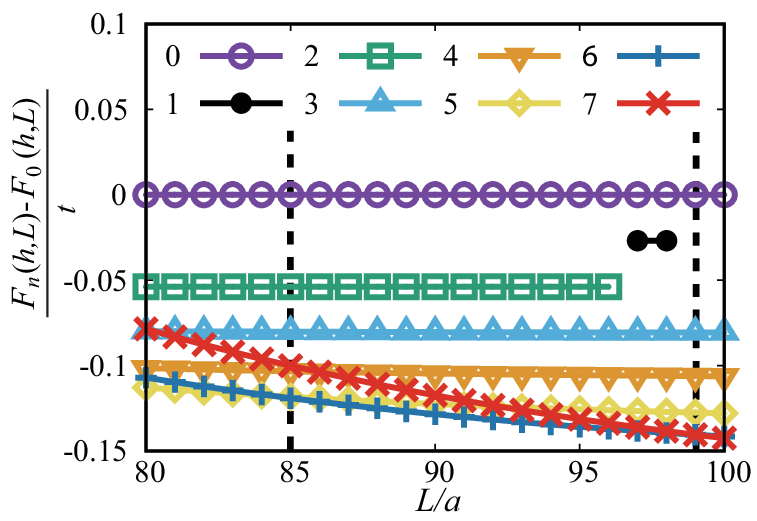}
    \includegraphics[width=\linewidth]{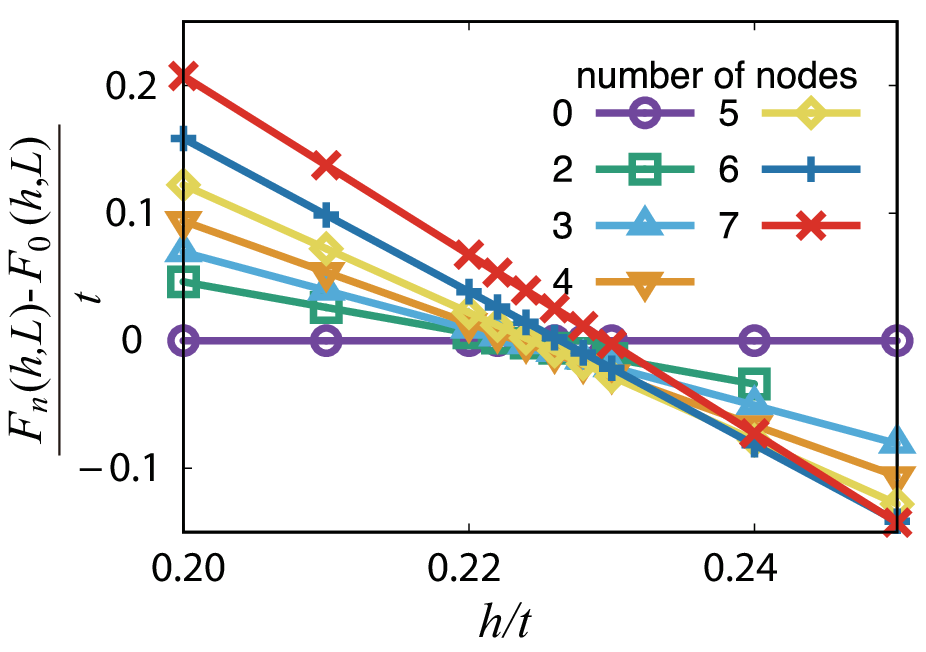}
    \caption{Free energy difference from the zero-node solution \textit{vs} $L$ (upper panel) and  $h$ (lower panel) for the ground state and metastable solutions. In the upper panel we have set $h=0.25$ and used the same parameters as in Fig.~\ref{Delta1}. In the lower panel, we have set $L/a=80$.  
In the upper panel, we can see that the number of nodes of the ground state changes from $5$ to $6$ at $L=85$ and $6$ to $7$ at $L=99$. The transition points are indicated by vertical lines. In the lower panel, we can also observe an increase in the number of nodes of the ground state as the magnetic field increases.}
    \label{freeEsizedep}
\end{figure}

The transition can be seen also by plotting the free energy of each solution as a function of $L$ and $h$. This is shown more extensively in Fig.\ \ref{freeEsizedep}, where we plot the free energy $F_n$ (normalized by the free energy of the zero-node solution $F_0$) of each configuration with $n$ nodes, including the metastable ones, as a function of the system size $L$ and magnetic field $h$. The parameters are chosen as before and we display the region of $L$ and $h$ where the transition between different solutions is more evident. 
In the upper panel, the free energy difference from the zero-node solution for each system size are plotted for $h/t=0.25$. The range of display is chosen to illustrate that the lowest energy state switches from $n=5$ to $n=6$ at $L/a=85$ and to $n=7$ at $L/a=99$ as the size $L$ is increased (transition points are indicated by vertical lines). As we have explained, these transitions are not continuous. The lower panel shows the free energy for the lowest energy state as a function of $h$ with fixed system size $L/a=80$.

\begin{figure}
    \centering
    \includegraphics[width=\linewidth]{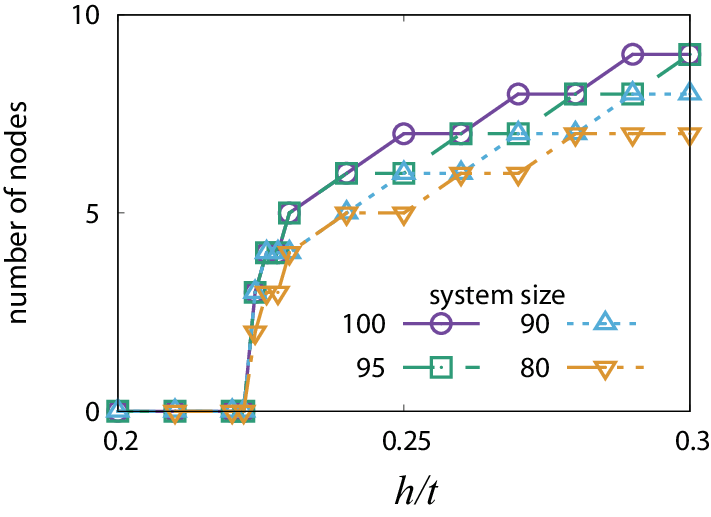}
    \caption{ 
    Number of nodes for the lowest energy state as functions of the magnetic field for various system sizes. 
    }
    \label{nodesvVSmag}
\end{figure}

The number of nodes of the ground state is more clearly visualized in Fig.~\ref{nodesvVSmag} where, for the same range of parameters as in Fig.~\ref{freeEsizedep}, we plot the number of nodes of the ground state vs the magnetic field for different sizes of the system.


\section{Casimir force} 
Before turning to the illustration of the force, we should more precisely discuss the connection between the Casimir effect and its pseudo counterpart we consider here. As we have seen, in the present case, the free-energy is made up of three terms: a semiclassical ground-state contribution $E_{GS}$, a pseudo-Casimir term $E_{PC} $ and a thermal correction $E_{T}$. We can focus for the time being on the first two terms. Let us consider $E_{GS}$. Because of the presence of interactions and non-periodic boundary conditions, the ``vacuum'' ground state of the model is inhomogeneous and it is determined by minimizing the whole effective action: in other words, the fluctuations that in the case of the electromagnetic field occur on top of real classical vacuum (this is essentially because the underlying quantum field theory is non-interacting), here take place on top of the ``vacuum'' of the theory that is a spatially inhomogeneous ground state. Because of this, the spectrum of the ``zero-point'' fluctuations depends in the present case on the ground state itself. This is a complication because minimization of the energy, which depends on the spectrum and thus on the ``vacuum'', has to be done self-consistently. In this sense, $E_{GS}$ is a semi-classical contribution that comes from the minimization of the free energy including quantum effects. The pseudo-Casimir contribution $E_{PC}$ comes from the fluctuations on top of the above ``vacuum'' ground state and has the typical Casimir form of a sum over the spectrum of the ``zero-point'' energies. These, differently from the usual Casimir effect where the deformations of the spectrum happen on top of the electromagnetic vacuum, here occur on top a soliton-like configuration. The thermal correction $E_{T}$ also depends on the spectrum and on the ``vacuum'' ground state, but its computation does not pose any problem, since, as always in quantum field theory, thermal contributions do not affect the divergences of the theory. 

\begin{figure}
    \centering
    \includegraphics[width=\linewidth]{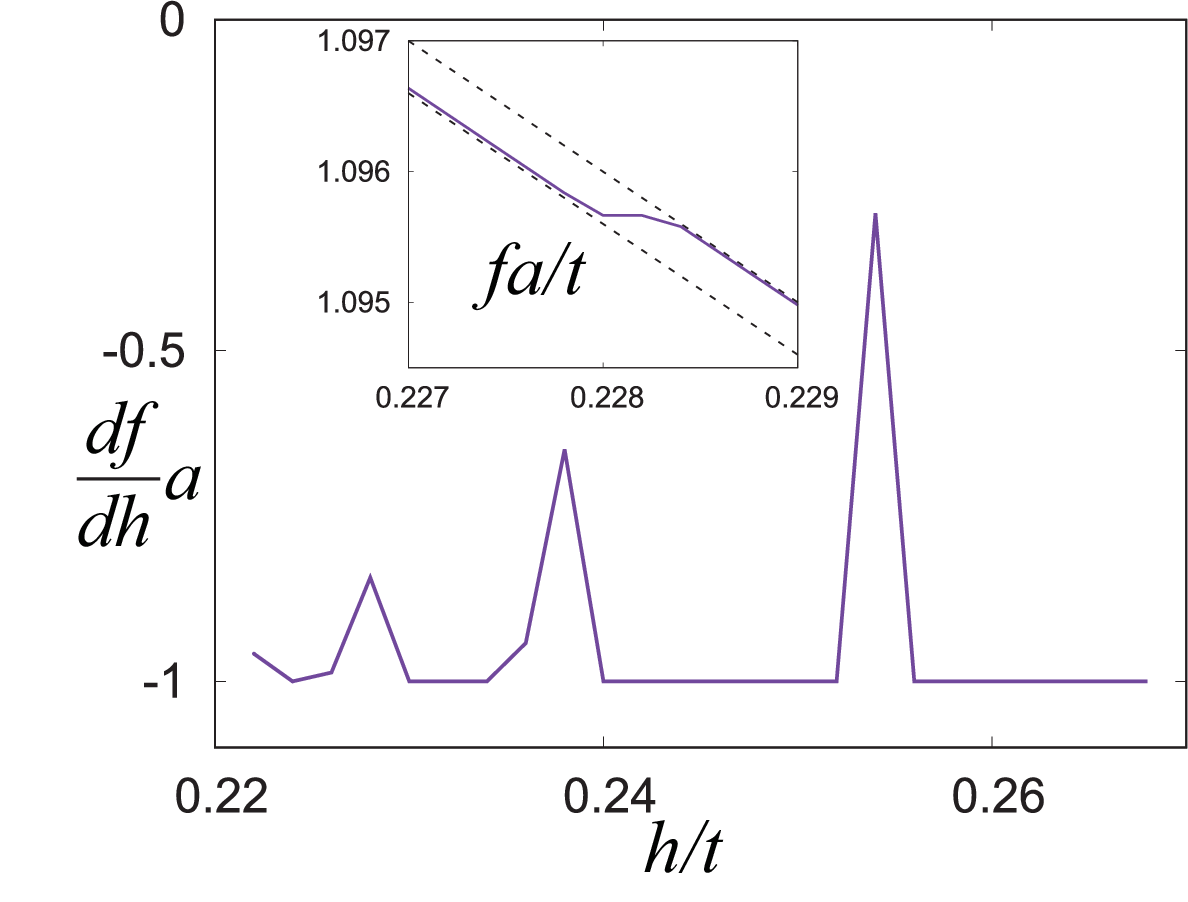}
    \includegraphics[width=\linewidth]{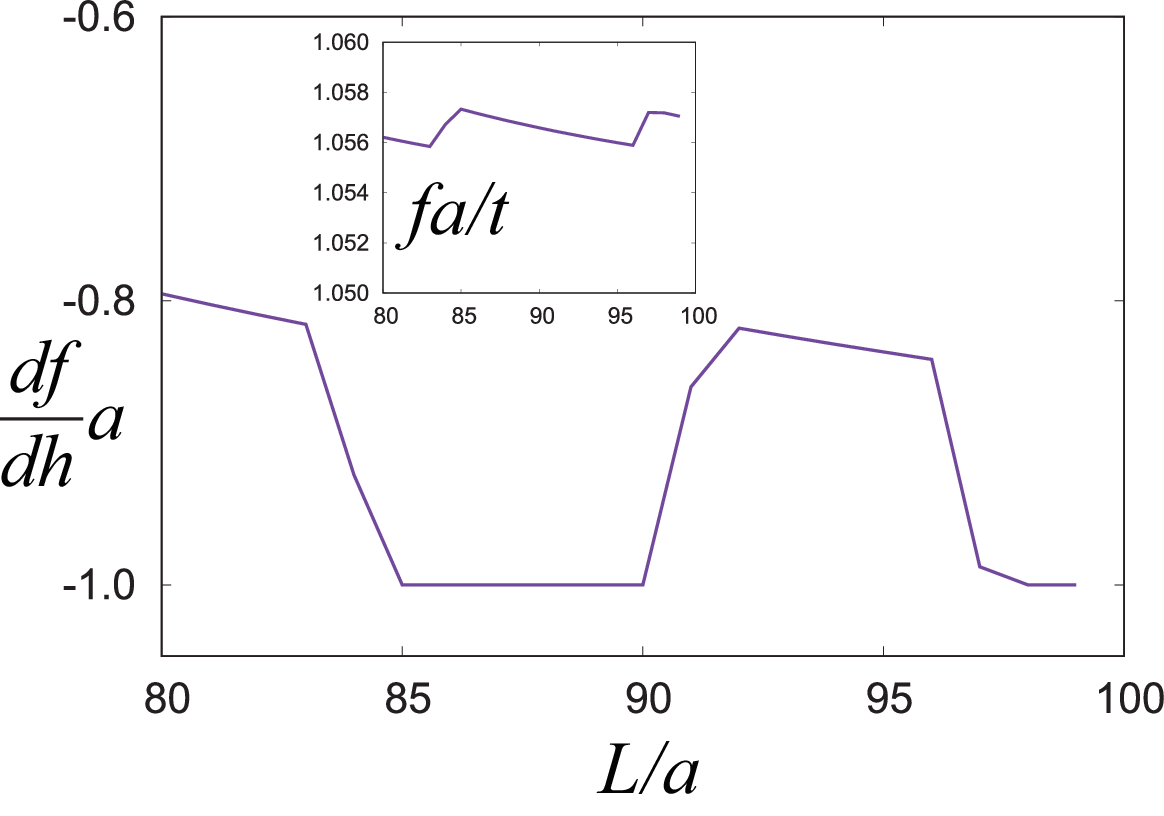}
       \caption{Casimir force behavior near the critical points. (Upper panel) Dependence of the derivative of the Casimir force $df/dh$ on $h$ for $L/a=80$. (Lower panel) Dependence of $df/dh$ on $L/a=80$ for $h/t=0.27$. The inset show the behavior of $f$ on $h/t$ (upper panel) and $L/a$ (lower panel).}
    \label{casimirlength}
\end{figure}

With the above clarifications in mind, we can try to gain some intuition on the Casimir force. As we have seen from the numerical computations, increasing either the size or the magnetic field changes the ``vacuum'' of the system and with it the spectrum of the zero-point fluctuations. Since these changes are discontinuous, then discontinuities should occur also in the pseudo-Casimir force as the external parameters, $g$, $L$ and $h$, are varied. From the free energy, we can obtain the pseudo-Casimir force by calculating $f\equiv -{\Delta F}/{\Delta L}$. 
The numerical calculations, at this stage do not present any particular difficulty since we have taken a discrete approach to the problem and thus the results are regularized by effect of discretization.
In Fig.\ \ref{casimirlength} we plot the derivative of the force $d f / dh$ as functions of $h$ (upper panel) and $L$ (lower panel). Parameters are chosen as in the previous figures. The occurrence of a jump in the force (which is shown in the inset for both plots) corresponds to the phase transition points (discontinuities in the derivative) that occur when the ground state jumps to a new solution with a larger number of nodes. In the present model, the Zeeman term $h\sigma$ contributes to the difference $F(L) - F(L-1)$ with an amount $\sim h\times L$ leading to $df/dh\sim -1$ away from the critical point. Indeed, as shown in Fig.\ \ref{casimirlength}, we find $df/dh\sim -1$ except for the vicinity of the phase transition points. The sudden decrease of $df/dh$ means that the Casimir force becomes less sensitive against the change of $h$ near the transition points.

\section{Conclusions}
 
In this paper, we have studied a simple system of interacting fermions confined to 1 spatial dimension 
in the presence of a constant, planar magnetic field, which appears as a shift in the chemical potential of the system. We have modeled the system and the associated superconducting transition by means of a Bogoliubov-de Gennes equation discretized over an interval, which we have solved numerically in a self-consistent fashion. First of all, we have observed that the magnetic field shifts the ground state discontinuously from solutions with a smaller to larger number of nodes when excess spin can be accommodated at nodal points in an energetically more convenient way. More precisely, we have found that the number of nodes of the inhomogeneous ground state of the system changes, as the size and magnetic field are varied, producing first-order transitions between the different solutions. 
The changes in the ground state have an influence on the Casimir energy of the system, which we have computed from the free energy. The numerical calculations indicate that, in the vicinity of the transition points, the magnetic field dependence causes a singular behavior where $df/dh$ drastically changes as a function of $h$. 
The present work can be extended to other boundary conditions and to other quantum field theory models 
\cite{Flachi:2017xat,Nitta:2017uog,Betti:2017zcm,Bolognesi:2018njt,Nitta:2018lnn,Nitta:2018yen,Yoshii:2019yln}.
The Casimir jump due to nodes would be an universal phenomenon in nontrivially interacting systems, in contrast to widely-studied Casimir effects of noninteracting fields.

\section*{Acknowledgments} 
The support of the Japanese Society for the Promotion of Science [Grant-in-Aid for Scientific Research, KAKENHI, Grant No. 21K03540 (AF), JP22H01221 (MN), JP24K06974 (ST), 19K14616 and 20H01838 (RY).] and of  the WPI program ``Sustainability with Knotted Chiral Meta Matter (SKCM$^2$)'' at Hiroshima University (MN, RY).




\end{document}